\documentclass[10pt,aps,prb,twocolumn,showpacs,superscriptaddress]{revtex4-1}%\documentclass[aps,prb,preprint,groupedaddress]{revtex4-1}
\usepackage[T1]{fontenc}
\usepackage{graphicx}

\begin{document}

%\preprint{}

\title{Thermodynamic and thermoelectric properties of {(Ga,Mn)As} and related compounds}


\author{Cezary \'Sliwa}
\email{sliwa@ifpan.edu.pl}
%\thanks{}
\affiliation{Institute of Physics, Polish Academy of Science, al.~Lotnik\'ow 32/46, PL-02-668 Warszawa, Poland}

\author{Tomasz Dietl}
\affiliation{Institute of Physics, Polish Academy of Science, al.~Lotnik\'ow 32/46, PL-02-668 Warszawa, Poland}
\affiliation{Institute of Theoretical Physics, Faculty of  Physics, University of Warsaw, PL-00-681 Warszawa, Poland}


\date{\today}

\begin{abstract}
Various experimental results providing information on thermodynamic density of states in (Ga,Mn)As are analyzed theoretically assuming that holes occupy GaAs-like valence bands. Allowing for Gaussian fluctuations of magnetization, the employed model describes correctly a critical behavior of magnetic specific heat found experimentally in {(Ga,Mn)As} near the Curie temperature $T_{\text{C}}$ [S. Yuldashev {\em et al.}, Appl. Phys. Express {\bf 3}, 073005 (2010)]. The magnitudes of room temperature thermoelectric power, as measured for GaAs:Be and (Ga,Mn)As [M. A. Mayer {\em et al.}, Phys. Rev. B {\bf 81}, 045205 (2010)], are consistent with the model for the expected energy dependencies of the hole mobility. The same approach describes also temperature variations of conductance specific to the Anderson-Mott localization, found for various dimensionality (Ga,Mn)As nanostructures at subkelvin temperatures [D. Neumaier {\em et al.}, Phys. Rev. Lett. {\bf 103}, 087203 (2009)]. We conclude that the examined phenomena do not provide evidence for an enhancement of density of states by the presence of an impurity band at the Fermi energy in ferromagnetic (Ga,Mn)As. Furthermore, we provide for (Ga,Mn)As expected values of both electronic specific heat at low temperatures $T \ll T_{\text{C}}$ and magnetization as a function of the magnetic field at $T_{\text{C}}$.

\end{abstract}

%\keywords{}

\maketitle

\section{Introduction}

Extensive studies of hole-controlled ferromagnetic semiconductors, such as (Ga,Mn)As and $p$-(Cd,Mn)Te, have allowed to demonstrate a variety of functionalities specific to these systems.\cite{Ohno:2010_NM,Dietl:2010_NM} At the same time, however, results of various experiments have faced us with a number of challenges suggesting that the understanding of these materials is by far not satisfactory. For instance, recent studies of (Ga,Mn)As using two kinds of tunneling spectroscopy have lead to two entirely contradictory pictures of electronic states in the vicinity of the Fermi energy $E_{\text{F}}$. According to scanning tunneling microscopy,\cite{Richardella:2010_S} and in agreement with the Altshuler and Aronov\cite{Altshuler:1985_B} and Finkelstein\cite{Finkelstein:1990_SSR} description of the Anderson-Mott localization for the spin polarized band carriers,\cite{Wojtowicz:1986_PRL} the one-electron density of states (DOS) attains a minimum at $E_{\text{F}}$ and its depression extends over an energy range of the order of the momentum relaxation rate, $\hbar/\tau \sim 100$~meV in (Ga,Mn)As.\cite{Jungwirth:2007_PRB} In contrast to one-electron DOS probed in tunneling experiments, the DOS for charge excitations as well as thermodynamic DOS, $\rho_{\text{F}} =\partial p/\partial E_{\text{F}}$, are only weakly renormalized by carrier correlation and disorder in doped semiconductors on the metal side of the Anderson-Mott transition, where the ratio of the inter-particle distance to an effective Bohr radius is relatively small $r_s = (4\pi p /3)^{-1/3}/a_B^* \lesssim 2.4$.

A rather different picture emerges from resonant tunneling spectroscopy of (Ga,Mn)As quantum wells.\cite{Ohya:2010_PRL,Ohya:2011_NP}  According to the  interpretation of the accumulated findings,  the Fermi level resides in a narrow impurity band located $\sim 50$~meV above the edge of hole subbands. These subbands are well resolved by resonant tunneling spectroscopy and, thus, virtually unaffected by disorder.\cite{Ohya:2010_PRL,Ohya:2011_NP}

The impurity band model implies an enormous thermodynamic DOS of carriers at the Fermi level. An unusually large effective mass emerges also within some interpretations of the optical\cite{Burch:2008_JMM} and transport\cite{Alberi:2008_PRB} properties of (Ga,Mn)As.
Historically, large magnitudes of effective masses in,
for example, heavy-fermion systems\cite{Stewart:1984_RMP} and Kondo alloys\cite{Kondo:1969_B,Heeger:1969_B} were discovered by studies of thermodynamic and thermoelectric effects. Therefore, it is of particular importance to describe theoretically those properties of (Ga,Mn)As, which provide information on thermodynamic DOS. Since this DOS, in contrast to one-electron DOS probed in tunneling experiments, does not exhibit a Coulomb anomaly at $E_{\text{F}}$,\cite{Altshuler:1985_B} it should show a significant enhancement if relevant electronic states are indeed characterized by a large effective mass.

In this paper, various experimental results providing information on thermodynamic DOS in (Ga,Mn)As are analyzed theoretically assuming that holes occupy GaAs-like valence bands, described here by the six band $k \cdot p$ model with effects of the $p$-$d$ exchange interaction included within the molecular field approximation.\cite{Dietl:2000_S,Dietl:2001_PRB} We show that the employed model, taking into account Gaussian fluctuations of magnetization, describes correctly a critical behavior of magnetic specific heat found experimentally in (Ga,Mn)As around the Curie temperature $T_{\text{C}}$.\cite{Yuldashev:2010_APEX} We also provide the expected field dependence of magnetization at $T_{\text{C}}$ within the mean-field approximation. Furthermore, we show that the magnitudes of room temperature thermoelectric power, as measured for GaAs:Be and (Ga,Mn)As,\cite{Mayer:2010_PRB} are consistent with our model for the expected energy dependencies of the hole mobility. We then turn to low temperatures $T\ll T_{\text{C}}$ and present computed values of the Sommerfeld coefficient $\gamma$, describing electronic specific heat, and magnetization as a function of the magnetic field at $T_{\text{C}}$. We also employ the same approach to discuss temperature variations of conductance specific to the Anderson-Mott localization, determined experimentally for various dimensionality (Ga,Mn)As nanostructures at subkelvin temperatures.\cite{Neumaier:2009_PRL} We conclude that the examined phenomena do not provide any evidence for an enhancement of thermodynamic DOS by the presence of an impurity band at $E_{\text{F}}$ in ferromagnetic (Ga,Mn)As.
\section{Critical behavior of specific heat}
\label{sec:gf}

Recently, a critical behavior of specific heat in Ga$_{1-x}$Mn$_x$As was resolved experimentally for two samples with the Mn concentration $x = 1.6$ and 2.6\%, which showed insulating and metallic behavior, respectively.\cite{Yuldashev:2010_APEX}

In order to describe these findings we employ the Ginzburg-Landau approach taking into account critical fluctuations in the Gaussian approximation.\cite{Ma:2000_B} We specify by $\sigma_i(x)$, $i = 1, 2, \ldots, n$,
a local magnitude of an $n$-dimensional order parameter (e.~g., a magnitude of local spin density) around $x$ in a $d$ dimensional block. The block Hamiltonian (the free energy functional) $H[\sigma(x)]$ at temperature $T$ and in the magnetic field $h$ can then be written as an expansion in powers of $\sigma(x)$ and $\nabla\sigma(x)$,\cite{Ma:2000_B}
\begin{eqnarray}
  H[\sigma(x)]/k_{\text{B}} T & = & \int \mathrm{d}^d x \,  \bigl[
    a_0 + a_2 \, \sigma^2 + {} \bigr. \label{eq:gl}\\
  & & \qquad \bigl.
    {} + a_4 \left(\sigma^2\right)^2 + c \left( \nabla \sigma \right)^2 - h \cdot \sigma \bigr],
  \nonumber
\end{eqnarray}
where
\begin{eqnarray}
  \sigma^2 & = & \sum_{i = 1}^n \left(\sigma_i(x)\right)^2, \\
	\left( \nabla \sigma \right)^2 & = & \sum_{\alpha = 1}^d \sum_{i = 1}^n \left(
	  \frac{\partial \sigma_i}{\partial x_\alpha} \right)^2.
\end{eqnarray}
The temperature dependent coefficients $a_k$ ($k = 0, 2, 4$) and~$c$ describe the free energy cost associated with a change in the magnitude and in the local direction of the order parameter, respectively.

Within the Gaussian approximation and taking into account only terms that are singular on approaching the ordering temperature $T_{\text{C}}$, the specific heat assumes a critical behavior given by\cite{Ma:2000_B} $C = C^{{\pm}} t^{d/2 - 2}$, where $C^{{+}} = n C_0$ and
$C^{{-}} = 2^{d/2} C_0$ apply to the $T > T_{\text{C}}$ and $T < T_{\text{C}}$ cases, respectively; $t =|T-T_{\text{C}}|/T_{\text{C}}$; and
\begin{equation}
  \frac{C_0}{k_{\text{B}}} = \frac12 (2\pi)^{-d} (a_2' T_{\text{C}}/c)^{\frac{d}{2}} \int \mathrm{d}^d k' \, (1+k'^2)^{-2},
\end{equation}
where $a_2'$ defined by $a_2 = a_2' (T-T_{\text{C}})$ shows no singular dependence on temperature around $T_{\text{C}}$.

In order to determine the magnitudes of $a_2'$ and $c$ for (Ga,Mn)As we note that, within the $p$-$d$ Zener model,\cite{Dietl:2000_S,Jungwirth:2006_RMP,Dietl:2007_B} the free energy functional consists of two contributions. The first describes the free energy of localized spins of magnitude~$S$ in the absence of carriers, whereas the second is the free energy of the Fermi liquid of holes in the valence band in the molecular field of a prescribed configuration of the localized spins (for such a description to be possible it is required that the dynamics of the localized spins are much slower than that of the carriers). The form of particular contributions to the Ginzburg-Landau free energy functional was determined\cite{Dietl:2000_S,Dietl:2007_B,Werpachowska:2010_PRB}   employing magnetization of localized spins $M(x) \equiv \sigma(x)M_{\text{Sat}}/S$ as an order parameter, where the saturation magnetization is related to the magnetic moment $g\mu_{\text{B}}S$ and the effective concentration of localized spins $N_0 x_{\text{eff}}$ in a standard way $M_{\text{Sat}} = g\mu_{\text{B}}SN_0 x_{\text{eff}}$, where the $g$-factor $g \approx 2.0$.  Neglecting interactions between localized spins in the absence of carriers, and assuming that the carrier liquid is strongly degenerate,\cite{Dietl:2000_S,Dietl:2007_B} we obtain
%for temperatures close to $T_{\text{C}}$
\begin{equation}
  a_2' T_{\text{C}} = \frac{3x_{\text{eff}} N_0}{2S(S+1)}
\end{equation}
and, in terms of the magnetic stiffness $A$,\cite{Werpachowska:2010_PRB,Konig:2001_PRB}
\begin{equation}
  c  = \lim_{M \to 0} \frac{A(M)M_{\text{Sat}}^2}{S^2M^2 k_{\text{B}} T_{\text{C}}} \equiv
    \frac{(N_0 \beta x_{\text{eff}})^2}{k_{\text{B}} T_{\text{C}}} \frac{B}{2},
\end{equation}
where $B$ is a property of the electronic subsystem, $N_0$~is the cation concentration and $\beta$ is the $p$-$d$ exchange constant.

In Ref.~\onlinecite{Yuldashev:2010_APEX}, experimental data were presented for the specific heat of two (Ga,Mn)As films.  According to x-ray diffraction measurements and x-ray microanalysis, the Mn concentration $x$ is 1.6\% in the sample~A and 2.6\% in the sample~B, with $T_{\text{C}}$ of $40$
and $52 \,\mathrm{K}$, respectively. Due to relatively low $x$ values, these samples are expected to be close to the metal-insulator transition (MIT). This expectation is confirmed by temperature dependencies of resistance, which  indicate that samples A and B are on the insulating and metallic side of the MIT, respectively.\cite{Yuldashev:2010_APEX} According to magnitudes of the Hall resistance $\rho_{\text{H}}$ at room temperature, the hole concentrations are $2.7 \times 10^{19}$ and $4.5 \times 10^{19}$ cm$^{-3}$ for samples A and B, respectively. Since, however, $\rho_{\text{H}}$ tends to diverge near the MIT\cite{Jaroszynski:1992_PB} (if measured at temperatures below the impurity binding energy, $\sim 1000 \, \mathrm{K}$ in GaAs:Mn), it leads to underestimated values of the hole concentrations. Furthermore, owing to critical fluctuations in the local DOS near the MIT,\cite{Dietl:2000_S,Dietl:2008_JPSJ,Richardella:2010_S}  only a part of the sample volume is ferromagnetic.\cite{Dietl:2000_S,Dietl:2008_JPSJ}  This means that the apparent values of low-temperature spontaneous magnetization are smaller than the expected magnitudes of saturation magnetization $M_{\text{Sat}} = g\mu_{\text{B}}SN_0x_{\text{eff}}$, as observed.\cite{Mayer:2010_PRB,Myers:2006_PRB,Sawicki:2010_NP}

Assuming that $x_{\text{eff}} < x$ only due to the presence of interstitial Mn, we have $x_{\text{eff}} = x - 2 x_{\mathrm{i}}$, where
$x$ and~$x_{\mathrm{i}}$ are the total and interstitial Mn contents. Similarly, the hole concentration is $p = N_0 (x - 3 x_{\mathrm{i}})$.\cite{Stefanowicz:2010_PRB} Using these equations and the mean-field theory,\cite{Dietl:2000_S} we determine the values
of $x_{\mathrm{eff}}$ and $p$, which reproduce the experimental values of $T_{\text{C}}$. We find
$x_{\text{eff}} = 1.6$\% and $p =3.5 \times 10^{20} \, \mathrm{cm^{-3}}$
for the sample~A, whereas for the sample~B the values are 2.0\% and
$3.8 \times 10^{20} \, \mathrm{cm^{-3}}$, respectively. As expected, the magnitudes of the hole concentrations obtained in this way are larger than the ones obtained from the Hall resistance, as quoted above. Similarly, the measured values of spontaneous  magnetization\cite{Yuldashev:2010_APEX} are by a factor 4.7 and~4.0 smaller than $M_{\text{Sat}}$ calculated for the values of $x_{\mathrm{eff}}$ for samples A and B, respectively. We note that the contribution of holes' magnetic moment to experimentally available magnetization may account for its reduction by less than 20\%.\cite{Sliwa:2006_PRB}

\begin{figure}
  \begin{center}
        \includegraphics[width=0.92\columnwidth]{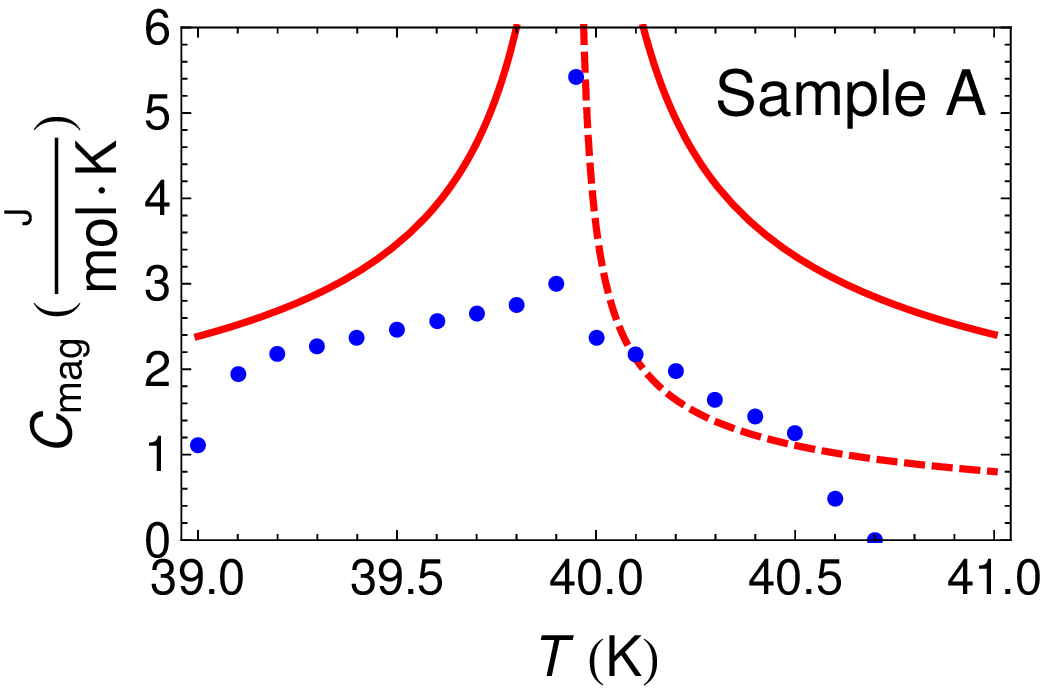}
        \includegraphics[width=0.944\columnwidth]{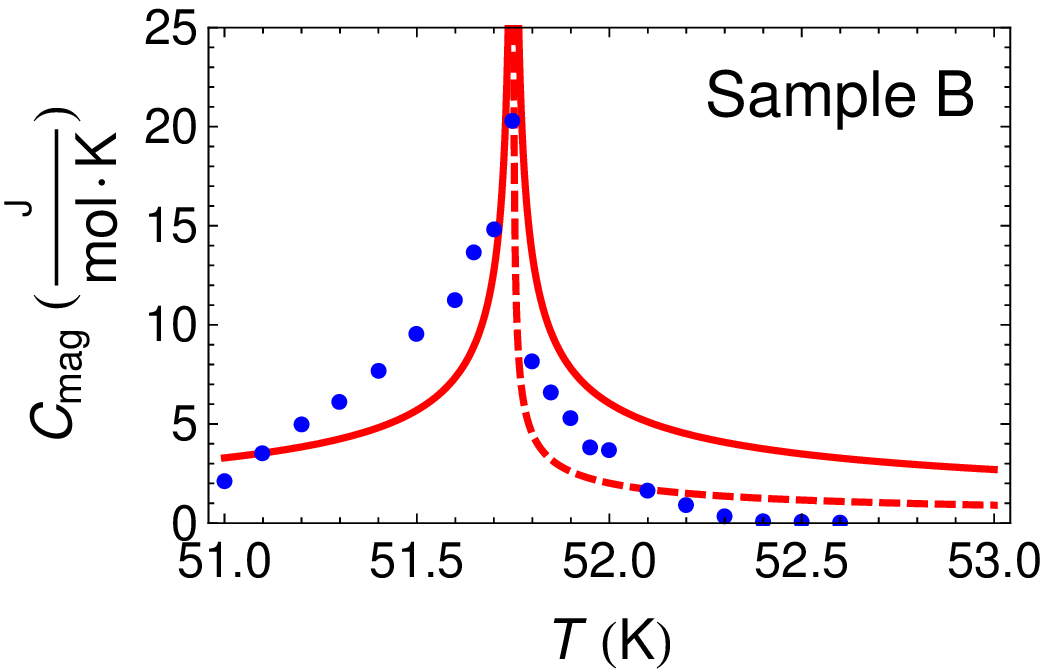}
  \end{center}
  \caption{(Color online) Theoretical temperature dependence of the magnetic specific heat calculated with no adjustable parameters for the sample~A (upper panel) and the sample~B (lower panel).
    The solid and dashed lines correspond  to the Heisenberg ($n = 3$) and Ising ($n = 1$) models, respectively.
    Dots represent experimental data (after Yuldashev \textit{et al.}\cite{Yuldashev:2010_APEX}).}
  \label{fig:A}
\end{figure}

We use the above sample parameters as well as GaAs Luttinger parameters  and $N_0 \beta = -1.2 \, \mathrm{eV}$.\cite{Dietl:2000_S,Dietl:2001_PRB}
%, we obtain $A_F \rho_s/4 = 1.1 \, \mathrm{eV^{-1} nm^{-3}}$ and $1.2 \, \mathrm{eV^{-1} nm^{-3}}$ for samples A and~B, respectively.
The values of~$B$ have been determined
within the six-band $k\cdot p$-model,\cite{Werpachowska:2010_PRB,Konig:2001_PRB}
in which we have neglected $E^{{+}{+}}_{\mathbf{k}}$. The numerical value is
$B = 0.17 \, \mathrm{eV^{-1} nm^{-1}}$ for both samples. These numbers yield
$C_0 = 0.13 \, \mathrm{J/(mol\cdot K)}$ for the sample~A
and $C_0 = 0.14 \, \mathrm{J/(mol\cdot K)}$ for the sample~B.
The corresponding dependencies of the specific-heat on temperature are shown in Fig.\ \ref{fig:A}, where experimental data from Figs.\ 3 and~4 of Ref.~\onlinecite{Yuldashev:2010_APEX} are depicted with dots. We see that the theory describes quite reasonably the experimental values, particularly if one can assume that, owing to magnetic anisotropy, the system is effectively Ising-like. Actually, since near the MIT only a part of the spins contribute to the ferromagnetic order, the calculated magnitude of the specific heat constitutes an upper limit of the expected values.

\section{Critical behavior of magnetization}

\begin{figure}
  \begin{center}
        \includegraphics[width=0.95\columnwidth]{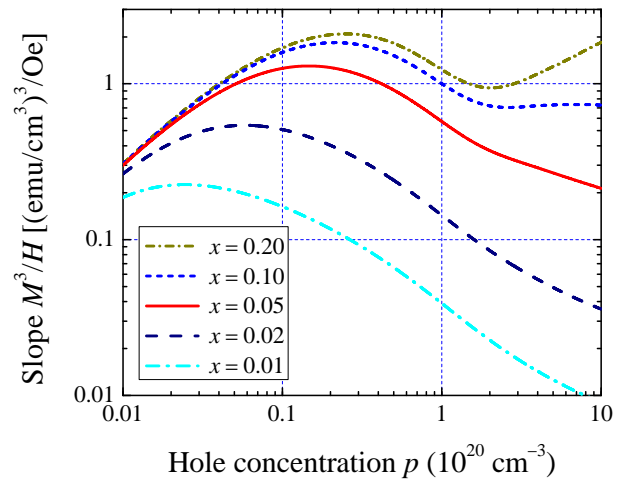}
  \end{center}
  \caption{(Color online) Computed values of the proportionality coefficient between $M(H)^3$ and $H$ for Mn magnetization $M$ in the magnetic field $H$ at $T_{\text{C}}$. The calculation has been performed within the $p$-$d$ Zener model for various effective Mn contents~$x$ in (Ga,Mn)As.}
  \label{fig:slope}
\end{figure}

\begin{figure}
  \begin{center}
        \includegraphics[width=0.95\columnwidth]{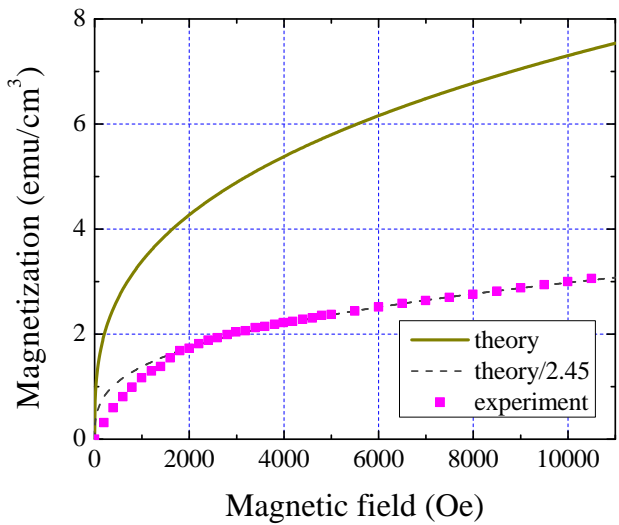}
  \end{center}
  \caption{(Color online) Comparison of theoretical (solid line) and experimental (symbols) field dependence of magnetization at $T_{\text{C}}$ for the sample~A of Yuldashev \emph{et al.}\cite{Yuldashev:2010_APEX} The dashed line shows theoretical values reduced by a factor of 2.45 to match experimental data.}
  \label{fig:M_vs_H}
\end{figure}

Within the Gaussian approximation, the temperature and field dependencies of magnetization $M(T,H)$ are given by the mean-field formula. Since theoretically expected values of spontaneous magnetization $M(T)$ were already  presented,\cite{Dietl:2001_PRB,Werpachowska:2010_PRB_b} we discuss here only $M(H)$ at $T = T_{\text{C}}$. This dependence is determined by the coefficient $a_4$ in Eq.~\ref{eq:gl} according to
\begin{equation}
  \bar\sigma = (h/4a_4)^{1/3}.
\end{equation}
Within the $p$-$d$ Zener model $a_4 = a_{4(S)} + a_{4(c)}$, where
\begin{equation}
  a_{4(S)} = \frac{9 (2 S^2 + 2 S + 1)}{40 S^3(S+1)^3} N_0 x_{\text{eff}} \approx
    6.2 \times 10^{-3} N_0 x_{\text{eff}},
\end{equation}
is the localized-spin contribution and $a_{4(c)}$,
the carrier contribution, is calculated numerically from the expansion of the carrier
free energy in the spin splitting parameter $B_G = A_{\text{F}}\beta M/(6g\mu_{\text{B}})$,
\begin{equation}
  F_c(B_G) = F_c(0) - p_2 B_G^2 + p_4 B_G^4,
\end{equation}
where $A_{\text{F}}$ is the Landau parameter describing the correlation-induced enhancement of the carrier spin susceptibility,  $A_{\text{F}} = 1.2$.\cite{Dietl:1997_PRB,Jungwirth:1999_PRB,Dietl:2000_S}, whereas the expansion coefficients  $p_i$ are determined by the band structure parameters and the hole concentration. We obtain
\begin{equation}
  M^3 = \frac{(g \mu_{\text{B}} N_0 x_{\text{eff}})^4}{4 k_{\text{B}} T_{\text{C}} (a_{4(S)} + a_{4(c)})} \mu_0 H,
\end{equation}
where
\begin{equation}
  k_{\text{B}} T_{\text{C}} = A_{\text{F}} \beta^2 N_0 x_{\text{eff}} S(S+1) p_2/54,
\end{equation}
and
\begin{equation}
  a_{4(c)} = A_{\text{F}} (\beta N_0 x_{\text{eff}}/6)^4 p_4 / k_{\text{B}} T.
\end{equation}
The computed magnitudes of the slope $M(H)^3/H$ at various hole concentrations and effective Mn contents are presented in Fig.~\ref{fig:slope}. The decrease of the slope with increasing hole concentration, seen in the plot for low~$x$, is due to an increase of $T_{\text{C}}$. The value for the sample~A is $0.039 \, {(\mathrm{emu}/\mathrm{cm}^{3})}^{3}/\mathrm{Oe}$, which corresponds to the solid line shown in Fig.~\ref{fig:M_vs_H} against the experimental data.\cite{Yuldashev:2010_APEX} As seen, the theoretical values of magnetization $M(H)$ reproduce satisfactorily the character of the field dependence observed experimentally. However, the absolute magnitudes of computed magnetization are by a factor of about 2.45 greater than the experimental ones. We attribute this discrepancy to a reduction of  the ferromagnetic phase volume by critical fluctuations in the local DOS in the vicinity of the MIT, as discussed in the previous section. In fact, this factor is even smaller than the ratio of saturated and low-temperature spontaneous magnetization, determined to be 4.7 for the sample in question, as discussed in the previous section.

\section{High temperature thermoelectric power}

In general, thermoelectric power contains diffusion and phonon drag contributions,\cite{Jedrzejczak:1976_PSS} as well as, in the ferromagnetic case, a magnon drag term.  The magnitudes of the drag terms scale with phonon and magnon relaxation times, so that they dominate at low temperatures, particularly in annealed (Ga,Mn)As samples,\cite{Osinniy:2001_APP,Osinniy:2008_SQO,Pu:2008_PRL} where a reduced concentration of Mn interstitials may suppress relevant scattering.

Awaiting for a quantitative theory of phonon and magnon scattering in (Ga,Mn)As, we limit our considerations to high temperatures, where the diffusion term is expected to dominate. For spherical bands with arbitrary dispersion, the diffusion thermopower for carriers with charge $+e$ is given by\cite{Zawadzki:1974_AP}
\begin{equation}
  S = \frac{k_{\text{B}}}{e} \frac{\left< \frac{E-E_{\text{F}}}{k_{\text{B}} T} \mu(E) \right>}{
    \left< \mu(E) \right>},
  \label{eq:tp}
\end{equation}
where $f_0$ in
\begin{equation}
  \left< A \right> = \int dE \, \left( - \frac{\partial f_0}{\partial E} \right) A(E) k^3(E)
\end{equation}
is the Fermi-Dirac distribution function and $k^3(E)$ describes the spherical band.
The mobility~$\mu$ depends on the hole energy $E$ with respect to the top of the corresponding valence band subband as
\begin{equation}
  \mu(E)= \mu_0 E^r,
\end{equation}
where the exponent~$r$ depends on the mechanism which limits the carrier mobility.
We have $r \approx -1/2$ for scattering by acoustic phonons and $r \approx 3/2$ for scattering
by ionized impurities.

\begin{figure}
  \begin{center}
        \includegraphics[width=0.95\columnwidth]{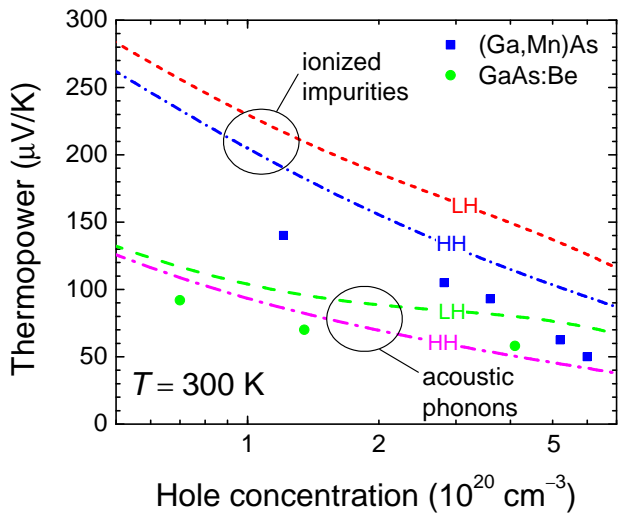}
  \end{center}
  \caption{(Color online) Room temperature thermoelectric power in (Ga,Mn)As and GaAs:Be as a function of hole density changed by irradiation with high energy Ne$^+$ ions (after Mayer \emph{et al.}\cite{Mayer:2010_PRB}). Lines are calculated using the standard six band model and GaAs parameters,\cite{Dietl:2001_PRB} and assuming that either ionized impurity or acoustic phonon scattering dominates. The actual value of thermopower $S$ should lie between lines obtained for heavy and light hole bands in each case.}
  \label{fig:thermpow}
\end{figure}

In the case of  a bulk GaAs-like semiconductor (zinc-blende structure, valence band maximum at the $\Gamma$~point), there are two kinds of carriers: heavy and light holes. Therefore, both the numerator and the denominator in Eq.~\ref{eq:tp}
are sums of the contributions from each of the subbands. Therefore, the resulting thermopower
coefficient is a weighted average of the coefficients calculated separately for each subband at the Fermi level, which is determined by the total concentration of holes distributed over both subbands.
Since the weights include the unknown parameter~$\mu_0$, which in general has a different value for each
subband, we cannot determine this average. However, we can consider~$S$ calculated for the heavy
and light holes subbands separately as the limiting cases, as the final result has to fall
within the range spanned by them.

The results of computations carried out for 300~K and employing the 6-band $k \cdot p$ model\cite{Dietl:2001_PRB}
are presented in Fig.~\ref{fig:thermpow}. The thermoelectric power~$S$ is shown
separately for the heavy and light holes. It is expected that the heavy holes dominate
for the case of scattering on ionized impurities, while the opposite is true
for scattering on acoustic phonons (we assume that the Fermi energy is small compared to the spin-orbit splitting of the valence band).

The theoretical data are compared to experimental values obtained at room temperature by Mayer \emph{et al.}\cite{Mayer:2010_PRB}
for a series of GaAs:Be and (Ga,Mn)As samples, in which hole density was changed by irradiation with high energy Ne$^+$ ions that introduce compensating donor defects.  We note that for these (Ga,Mn)As samples the content of substitutional Mn is $x = 0.045$, and the MIT occurs at the hole concentration $p_c\approx 3\times 10^{20}$~cm$^{-3}$,\cite{Mayer:2010_PRB} a value about two orders of  magnitude higher than $p_c$ for GaAs:Be. Nevertheless, the difference between the magnitude of $S$ in (Ga,Mn)As and GaAs:Be is only slight, which shows that the MIT has actually a little effect on the thermodynamic DOS, as could be expected for the Anderson-Mott localization.

We expect that acoustic phonon scattering, for which  $r = -1/2$, is relevant in GaAs:Be at 300~K. In contrast, in the case of (Ga,Mn)As, owing to the proximity to the MIT, the mobility is expected to increase with the carrier energy. At the same time, an additional compensation by interstitial Mn makes ionized impurity scattering, for which  $r = 3/2$, more significant. Thus, we can conclude that the data for (Ga,Mn)As fall in the range expected for hole transport in the GaAs valence band.

\section{Low temperature electronic specific heat}

\begin{figure}
  \begin{center}
        \includegraphics[width=0.95\columnwidth]{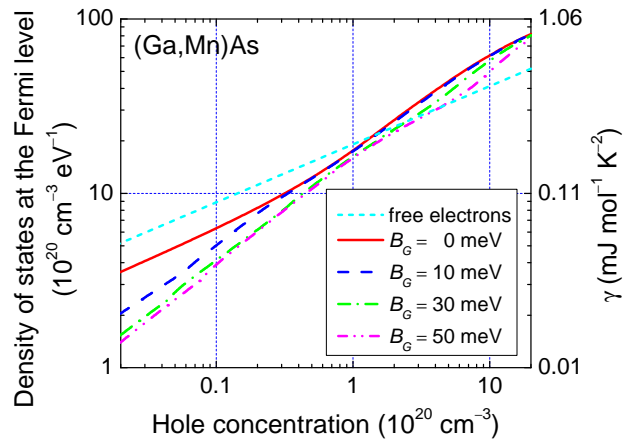}
  \end{center}
  \caption{(Color online) Thermodynamic density of states as a function of the hole concentration in (Ga,Mn)As for various values of the parameter $B_{\text{G}}$ characterizing spin splitting of the hole subbands.
  The corresponding values of the Sommerfeld electronic specific heat
    coefficient~$\gamma$ are shown on the right axis.}
  \label{fig:dos}
\end{figure}

In doped semiconductors on the metal side of the Anderson-Mott localization, the magnitude of the Fermi liquid parameter is relatively small $r_s = (4\pi p /3)^{-1/3}/a_B^*   \lesssim 2.4$, so that  Landau's renormalization of the specific heat and thermodynamic DOS by carrier-carrier interactions is of a minor quantitative importance. Hence, we compute the electronic specific heat in the low-temperature limit according to  $C_V = \gamma T$,
with the Sommerfeld constant
\begin{equation}
  \gamma = \frac{\pi^2}{3} k_{\text{B}}^2 \rho_{\text{F}},
\end{equation}
where $\rho_{\text{F}} =\partial p/\partial E_{\text{F}}$ is determined for holes in the valence band with no carrier-carrier interactions taken into account. We plot in Fig.~\ref{fig:dos} the dependence of $\rho_{\text{F}}$ and $\gamma$
on the hole concentration $p$ for (Ga,Mn)As with various values of the spin splitting parameter $B_{\text{G}} = A_{\text{F}}\beta M/(6g\mu_{\text{B}})$.
The numerical values for the samples A and~B of Yuldashev \textit{et al.}
are $0.37$ and~$0.39 \, \mathrm{mJ \, mol^{-1} \, K^{-2}}$, respectively, hence
the values of $\gamma T_{\text{C}}$ are small compared to the critical anomaly of the specific
heat, validating the results shown in Fig.~\ref{fig:A}.

\section{Low temperature conductivity}
Near the MIT, temperature and magnetic field dependencies of conductivity $\sigma(T,H)$ are determined by quantum phenomena specific to Anderson-Mott localization.\cite{Altshuler:1985_B,Lee:1985_RMP} These striking effects result from single-particle interferences of scattered waves and/or from scattering-driven interferences of carrier-carrier interaction amplitudes. It was suggested within this framework that the magnitude of conductivity changes in (Ga,Mn)As at low temperatures points to the value of DOS actually expected for the GaAs valence band.\cite{Dietl:2008_JPSJ} More recently, Neumaier \emph{et al.}\cite{Neumaier:2009_PRL} carried out comprehensive studies of conductance in various dimensionality structures of ferromagnetic (Ga,Mn)As at subkelvin temperatures.  Since the external magnetic field has no effect on $\sigma(T)$ in this regime,\cite{Neumaier:2008_PRB} the single-particle Anderson localization term, destroyed presumably by the demagnetizing field, does {\em not} contribute to $\sigma(T)$ in this ferromagnetic semiconductor below 1~K. At the same time, the study of $\sigma(T)$ at the dimensional crossover $\text{2D} \to \text{3D}$ allowed to determine the magnitude of the diffusion coefficient $D = \sigma/(e^2\rho_{\text{F}})$.\cite{Neumaier:2009_PRL} In this way the value of DOS at the Fermi level for charge excitations $\rho_{\text{F}}$ was determined and found to be slightly smaller than the one expected for the GaAs valence band.\cite{Dietl:2001_PRB,Neumaier:2009_PRL}

To supplement the above analysis, we consider $\sigma(T)$ within the universality class for which the transport proceeds in two subbands, whose splitting is much larger than $k_{\text{B}}T$ but much smaller than the Fermi energy. Furthermore, we make use of the value of the Landau parameter describing the correlation-induced enhancement of the carrier spin susceptibility and the Curie temperature $A_{\text{F}} = 1.2$.\cite{Dietl:1997_PRB,Jungwirth:1999_PRB,Dietl:2000_S} Our goal is to describe the magnitudes of parameters $a$ characterizing the rate of change of  $\sigma(T)$ displayed in Fig.~1 of Ref.~\onlinecite{Neumaier:2009_PRL} for various dimensionality systems.
In the 2D case $a =  F_{2D} e^2/(\pi h t)$, where according to theory developed for a simple isotropic band, in the presence of a sizable spin splitting, $F^{2D} = (1- F/4)\ln 10$, where $F = 2 (A_{\text{F}} -1)$.\cite{Altshuler:1985_B,Lee:1985_RMP} The theoretical value obtained for the sample thickness $t = 42$~nm, $a = 1.6\times 10^7 \, e^2/h\,\mbox{m}$ is seen to be in a good agreement with the experimental finding $1.8\times 10^7 \, e^2/h\,\mbox{m}$.\cite{Neumaier:2009_PRL}

We now apply the same procedure to the 1D and 3D cases, $d = 1$ and 3,\cite{Neumaier:2009_PRL} where $a \sim D^{1-d/2}$. Here, $F^{1D} = 1.37$ and $F^{3D} = 1.04$ are anticipated theoretically for $F = 0.4$.\cite{Altshuler:1985_B} For the experimental values  $\sigma = 6.95\times 10^8 \, e^2/h\,\mbox{m}$  and $3.95\times 10^8 \, e^2/h\,\mbox{m}$  as well as for $\rho_{\text{F}}= 1.98\times 10^{46}$~1/Jm$^3$ and $1.45\times 10^{46}$~1/Jm$^3$, as expected for disorder-free (Ga,Mn)As valence band at the hole concentrations in question,\cite{Dietl:2001_PRB,Neumaier:2009_PRL} we obtain: $a =  -3.1\times 10^7 \, (e^2/h)(\mbox{K}^{1/2}/\mbox{m})$ for $d = 1$ and $0.93\times 10^7 \, e^2/(h \, \mathrm{m} \, \mathrm{K}^{1/2})$ for $d = 3$,  which are close to the experimental values $a =  -2.5\times 10^7 \, (e^2/h)(\mbox{K}^{1/2}/\mbox{m})$ and $1.5\times 10^7 \, e^2/(h \, \mathrm{m} \, \mathrm{K}^{1/2})$, respectively, shown in Figs.~1b and~1d of Neumaier \emph{et al.}\cite{Neumaier:2009_PRL} As seen, this approach suggests a slightly higher DOS comparing to that expected for the GaAs-like valence band.

It is worth noting that other authors,\cite{Honolka:2007_PRB,Mitra:2010_PRB}  analyzing $\sigma(T)$ up to 4~K, found that $\sigma(T) = \sigma_0 +  AT^{\alpha}$, where $\alpha = 1/3$. This dependence was interpreted in terms of a renormalization group equation\cite{Altshuler:1985_B} applicable close to the MIT, where $ \sigma_0 <  AT^{\alpha}$ and then  $1/3 \lesssim \alpha \lesssim 1/2$ in the 3D case.\cite{Lee:1985_RMP,Belitz:1994_RMP} Furthermore, the apparent value of $\alpha$ can be reduced above 1~K by a cross-over to the regime, where the effect of scattering by magnetic excitations onto quantum corrections to conductivity becomes significant.\cite{Dietl:2008_JPSJ}

\section{Conclusions}

We have considered the thermodynamic and thermoelectric properties of (Ga,Mn)As including the critical behavior of specific heat and magnetization, high temperature thermoelectric power, low temperature electronic specific heat, and low temperature conductivity. The available experimental data\cite{Yuldashev:2010_APEX,Mayer:2010_PRB,Neumaier:2009_PRL} are consistent with the $p$-$d$ Zener model in which the carriers reside in a GaAs-like valence band. In particular, the critical behavior of specific heat\cite{Yuldashev:2010_APEX} can be reasonably well described assuming Gaussian fluctuations of magnetization. The magnitudes of the thermoelectric power at room temperature\cite{Mayer:2010_PRB} are consistent with the theoretical results for scattering mechanisms expected to limit the magnitude of hole mobility. The data for low-temperature electronic specific heat are provided in order to stimulate  corresponding experimental investigations. Finally, the magnitudes of density of states obtained assuming a valence band model are supported by the temperature dependence of conductivity at subkelvin temperatures, determined by disorder-modified carrier-carrier interaction effects, significant near the metal-insulator transition.\cite{Neumaier:2009_PRL} The proximity of this transition results also in a reduction of the volume occupied by spins contributing to the ferromagnetic order. We conclude that there is no experimental evidence for the enhanced density of states at the Fermi level, expected within the impurity-band models of ferromagnetism in (Ga,Mn)As. Instead, however, the properties of this ferromagnet are strongly affected by hole localization effects.

\section*{Acknowledgments}

This work was supported by the ``FunDMS'' Advanced Grant of the European Research Council within the Ideas 7th Framework Programme of the EC and within European Regional Development Fund through the grant Innovative Economy POIG.01.03.01-00-159/08 ``InTechFun''.

\bibliography{thermo}

\end{document}